\documentclass[twocolumn]{aastex63}

\usepackage{lineno}

\usepackage{amsmath}

\usepackage{calligra}

\usepackage{xspace}

\usepackage{color}

\DeclareMathAlphabet{\mathcalligra}{T1}{calligra}{m}{n}
\DeclareFontShape{T1}{calligra}{m}{n}{<->s*[2.2]callig15}{}
\newcommand{\scriptr}{\mathcalligra{r}\,}

\newcommand{\Tfrac}[2]{\left(\frac{#1}{#2}\right)}

\newcommand{\be}{\begin{enumerate}}
\newcommand{\ee}{\end{enumerate}}

\shorttitle{}
\shortauthors{Gallegos-Garcia et al.}

\begin{document}

\title{Angular Momentum Loss During Stable Mass Transfer onto a Compact object: the Effect of Mass Loss via Accretion Disk Winds } 
\author[0000-0003-0648-2402]{Monica Gallegos-Garcia}
\affiliation{Department of Physics and Astronomy, Northwestern University, 2145 Sheridan Road, Evanston, IL 60208, USA}
\affiliation{Center for Interdisciplinary Exploration and Research in Astrophysics (CIERA),1800 Sherman, Evanston, IL 60201, USA}

\author[0000-0003-2982-0005]{Jonatan Jacquemin-Ide}
\affiliation{Center for Interdisciplinary Exploration and Research in Astrophysics (CIERA),1800 Sherman, Evanston, IL 60201, USA}

\author[0000-0001-9236-5469]{Vicky Kalogera}
\affiliation{Department of Physics and Astronomy, Northwestern University, 2145 Sheridan Road, Evanston, IL 60208, USA}
\affiliation{Center for Interdisciplinary Exploration and Research in Astrophysics (CIERA),1800 Sherman, Evanston, IL 60201, USA}

\begin{abstract}
We use an analytic framework to calculate the evolution of binary orbits under a physically-motivated model that accounts for angular momentum loss associated with winds from an accretion disk around the compact-objected accretor. 
Our prescription considers wind mass ejection from the surface of an accretion disk, accounting for a radial mass-loss dependence across the disk surface.
We compare this to the standard prescription of angular momentum loss associated with isotropic mass loss from the vicinity of the accretor. 
The angular momentum loss from a disk wind is always larger. 
For mass ratios, $q$, between $2$--$10$, angular momentum loss via a disk wind is $\simeq3$--$40$ times greater than the standard prescription.
For the majority of mass ratios and disk properties, accounting for the disk wind can result in considerably smaller orbital separations compared to the standard formalism; the differences being $\simeq 60\%$ depending on how long the effect is integrated for.
We conclude that it is important to consider the effects of angular momentum loss from a disk wind when evolving binary orbits. 
\end{abstract}


\newpage

\section{Introduction}\label{sec:intro}
The fate of a binary system is largely determined by mass transfer (MT),  mass loss, and the associated angular momentum losses from the system. 
Various ``modes'' of MT have been formulated to describe different transfer scenarios and their subsequent mass and angular momentum loss \citep[for a review see,][]{vandenHeuvel_1994_interacting_binaries, Soberman1997, Postnov2014,2023Tauris_vandenHeuvel_book}.
MT, when a star overfills its Roche lobe (RL) and transfers mass onto its companion, is inevitable for close binaries.
Especially for the case of stable MT onto a compact object, a fraction of this mass is thought to be lost from the binary system entirely.
Mass loss associated with this scenario is typically described with the mode of isotropic re-emission. 
In this scenario, mass is re-emitted as a fast, isotropic wind that carries away with it the specific angular momentum of the accretor. 
In most cases, the spin angular momentum is typically negligible compared to the orbital angular momentum.
Other modes include mass loss from a circumbinary disk or the Jeans mode, which describes mass lost from the system due to stellar winds.

These modes of MT are implemented into large-scale binary population synthesis codes such as \texttt{COSMIC} \citep{Breivik2020}, \texttt{COMPAS} \citep{COMPAS_paper_2022}, and \texttt{POSYDON} \citep{POSYDON2022} as well detailed stellar and binary evolution codes such as \texttt{MESA} \citep{Paxton2011,Paxton2013,Paxton2015,Paxton2019}.
There is growing wealth of information from observations and detailed 3D accretion disk simulations that reveal that our assumptions for mass loss during stable mass transfer onto a compact object may need revision.

In observations of X-ray binaries, multiple authors have detected clear absorption line signatures  \citep{ponti_ubiquitous_2012,ponti_high_2016,tetarenko_strong_2018,munoz-darias_hard-state_2019,panizo-espinar_discovery_2022,mata_sanchez_ask_2023}. These absorption lines show a significant blue shift, consistent with outflowing material, with typical velocities of $100\,\,-\,\,2000\,\,\mathrm{km\,s^{-1}}$. Since the absorption lines are only present in high-inclination systems \citep{ponti_ubiquitous_2012}, they are indicative of equatorial winds, roughly perpendicular to the axis of rotation \citep{higginbottom_coronae_2015}. They are now commonly interpreted as disk winds ejected from the disk surface. However, the dynamical process that drives the winds remains unclear, as the resolution of the emission lines is not enough to distinguish between radiative or magnetic driving \citep{ponti_high_2016,chakravorty_magneto_2016,tomaru_thermal-radiative_2019,fukumura_modeling_2021,tomaru_what_2023,chakravorty_absorption_2023}. 

Although the X-ray absorption lines disappear during the hard state \citep{ponti_high_2016}, near-infrared absorption lines persist during the entire outburst of X-ray binaries \citep{sanchez-sierras_near-infrared_2020}. These results point to disk winds being an universal process of mass and angular momentum transport in X-ray binaries that might be independent of the intrinsic luminosity of the system, e.g., the Eddington ratio.

The density profile of the wind is directly linked to the accretion rate by mass conservation \citep{ferreira_magnetized_1995}. The steepness of the accretion rate as a function of the radius defines the mass ejection index, $\xi=\mathrm{d} \ln \dot{M}_{\rm disk}/\mathrm{d} \ln r$. When $\xi\simeq0.5$ the wind is heavily mass-loaded, while when $\xi<0.1$ the wind is tenuous \citep{jacquemin-ide_magnetically-driven_2019}. 
In the case of magnetic driving, the absorption lines provide steep constraints on the density profile of the disk wind and the mass ejection index.
Comparing the absorption lines from semi-analytical models of magnetically driven outflows \citep{ferreira_magnetized_1995} to observations one can constraint the value of the mass ejection index \citep{chakravorty_magneto_2016}.
Mass ejection index values of $\xi\simeq0.3-0.4$ are typically measured in such systems \citep{fukumura_modeling_2021}. Even though the mass ejection index has only been measured in black hole (BH) XRBs we expect it to be similar for neutron star XRBs due to the similar properties of their disk-winds \citep{ponti_high_2016}. 

The mass ejection index has also been measured in general relativistic magnetohydrodynamic (GRMHD) simulations of magnetically arrested accretion disks. \cite{mckinney_general_2012} measured $\xi\simeq 0.4$, consistent with constraints from X-ray binaries.

The assumptions made for stable mass transfer onto a compact object have consequences for a variety of systems such as cataclysmic variables, low-mass and high-mass X-ray binaries, pulsars, and binary compact mergers.
As described above, there is an abundance of evidence of mass loss via a disk wind. 
However, this mode of mass loss is currently not accounted for in standard binary evolution prescriptions.
In this paper we derive an analytic prescription that accounts for disk winds in the evolution of the binary's separation. 
We compare this to the standard prescription of isotropic re-emission.
In Section~\ref{sec:mass_loss} we describe how we model the disk wind. In Section~\ref{sec:angular_momentum_loss} we describe the geometry of the problem and how we calculate angular momentum loss from the disk wind. In Section~\ref{sec:results} we present our results focusing on the case of fully non-conservative MT and in Section~\ref{sec:conclusions} we present our conclusions and caveats. 
\\
\\

\section{Method}\label{sec:methods}

We assume that during stable MT all mass loss from the donor, $\dot{M}_{1}$, flows through the inner Lagrangian point and towards the accretor, forming an accretion disk.
We assume the material in the disk is instantaneously processed and a fraction $\beta$ of $\dot{M}_{1}$ is ejected from the system via the disk wind.
A fully non-conservative system is therefore described by $\beta=1$ and a system without any mass loss by $\beta=0$.
For the purposes of our analysis, we do not explicitly include specific details of mass loss, such as Eddington-limited MT or stellar winds, instead all non-conservative MT is modeled with $\beta$ and all mass loss occurs from the disk wind.
We assume the ejected mass is lost in the form of an azimuthally-symmetric wind originating from the surface of the accretion disk. 
We evolve the binary's orbital separation accounting for the orbital angular momentum loss from this wind.
In Section~\ref{sec:mass_loss} we describe the mass-loss profile of the disk wind. 
In Section~\ref{sec:angular_momentum_loss} we present the calculation for the total orbital angular momentum lost by the wind, and in Section~\ref{sec:evolve_binary} we describe how we evolve a binary through time.

\subsection{Wind mass loss from a disk} 
\label{sec:mass_loss}

We define the total rate of mass loss from the disk wind as $\dot{M}_{\rm w}$, which we assume is a fraction $\beta$ of the total mass loss from the donor, $\dot{M}_{\rm w} = \beta \dot{M}_{1}$.
The disk wind occurs from the surface of the disk and is described by,
\begin{equation}
    \dot{M}_{\mathrm{w}} = \iint 
  \dot{\sigma}_{\mathrm{w}} r \mathrm{d}r\mathrm{d}\phi,
  \label{eq:Mdot_wind}
\end{equation}
where $\dot{\sigma}_{\mathrm w}$ is the rate of wind mass loss per unit area on the disk, $r$ is the radial component within the disk, and $\phi$ is the azimuthal angle around the disk.
The bounds of integration are from 0 to $2\pi$ and from the inner radius of the disk wind, $r_{{\mathrm w}}$, to the outer radius of the disk, $r_{{\mathrm d}}$. 
The rate of wind mass loss per area on the disk is,
\begin{equation}
    \dot{\sigma}_{\rm w} = \frac{\xi \dot{M}_{\mathrm{w}}}{2 \pi r_{\mathrm{d}}^2} \left(\frac{r}{r_{\rm d}}\right)^{\xi-2},
    \label{eq:sigma_dot_wind_r}
\end{equation}
where $\xi$ is the mass ejection index. 
This leads to a cumulative mass ejection profile of $\int \dot{\sigma}_{\mathrm w}r\mathrm{d}r\propto r^{\xi}$.

We assume $r_{\mathrm{d}}$ is 75\% of the Roche lobe of the compact object, similar to \cite{King1997} and within the results of \cite{Paczynski1977_accretion_disks}. 
The dependence on the choice of $r_{\mathrm{d}}$ is briefly discussed in Section~\ref{sec:beta_r_disk}.
For the inner radius, we must distinguish the inner disk-wind radius $r_{{\mathrm w}}$ from the inner radius of the physical disk $r_{0}$. 
We expect $r_{0}$ to be of the order of a few the gravitational radii (assuming the compact object is a BH) and $r_{0}\ll r_{\mathrm{d}}$.
To eject a mass loss rate of $\dot{M}_{\rm w}$, the inner disk-wind radius may not extend to the physical edge of the accretion disk, $r_{{\mathrm w}}\geq r_{0}$.
For fully non-conservative accretion ($\beta = 1$) however, we can set $r_{{\mathrm w}} = r_{0} = 0$, as we do not expect inflowing material onto the accretor.

For partially conservative MT, the approximation of $r_{{\mathrm w}}=r_{0}=0$ is no longer valid. 
We must consider the inflow of mass {\emph{within}} the disk, which is determined by the radial profile of mass loss throughout the disk. 
A steep radial profile will sharply cut off the inflow of mass at larger disk radii compared to a shallow radial mass ejection profile. 
As a result, to set a fixed mass inflow rate we must solve for the inner radius of the disk wind given the mass inflow rate desired and the mass ejection profile.

The mass inflow rate at an arbitrary $r_{\mathrm{w}}$ is given by
\begin{equation}
    \dot{M}_{\mathrm{in}}(r_{\mathrm w}) = \dot{M}_{\mathrm{in}}(r_{\mathrm{d}}) - \int_{0}^{2\pi} \int_{r_{\rm w}}^{r_{\mathrm{d}}} 
  \dot{\sigma}_{\mathrm w} r\mathrm{d}r\mathrm{d}\phi,
  \label{eq:masscon}
\end{equation}
where $\dot{M}_{\mathrm{in}}(r_{\mathrm{d}})$ is the inflow rate at the outer radius of the disk, equal to $\dot{M}_{1}$, and the last term is the total mass ejected throughout the disk (Eqn.~\ref{eq:Mdot_wind}). 
With this we can write the accretion efficiency term as $\beta = 1 - \dot{M}_{\mathrm{in}}(r_{\mathrm w})/\dot{M}_{\mathrm{in}}(r_{\mathrm{d}})$ and reparameterize it as a function of inner disk-wind radius $r_{\mathrm w}$, the outer radius of the disk $r_{\mathrm d}$, and $\xi$,
\begin{equation}
    \beta = 1 - \left( \frac{r_{\mathrm w}}{r_{\mathrm{d}}} \right)^\xi.
    \label{eq:beta_r_in}
\end{equation}
We use Eqn.~\eqref{eq:beta_r_in} to solve for the inner disk-wind radius, $r_{\mathrm w}$, when $\beta$ does not equal 1.  
We enforce mass loss from the outer regions of the disk by adjusting $r_{\mathrm w}$ instead of allowing for extended mass loss in the inner regions of the disk. 
This is a conscious choice as mass loss from the outer regions of the disk maximizes the angular momentum loss, increasing the effect of the disk winds on the orbital evolution.
Furthermore, this choice is consistent with spectral modeling of XRB disk winds, where it is assumed that disk-winds extends up to the outer radius of the disk \citep{fukumura_modeling_2021,tomaru_what_2023}. Different radial extents for the disk wind will be considered in future work.

\subsection{Loss of angular momentum through disk winds} \label{sec:angular_momentum_loss}

\begin{figure}
\centering
\includegraphics[width=0.475
\textwidth]{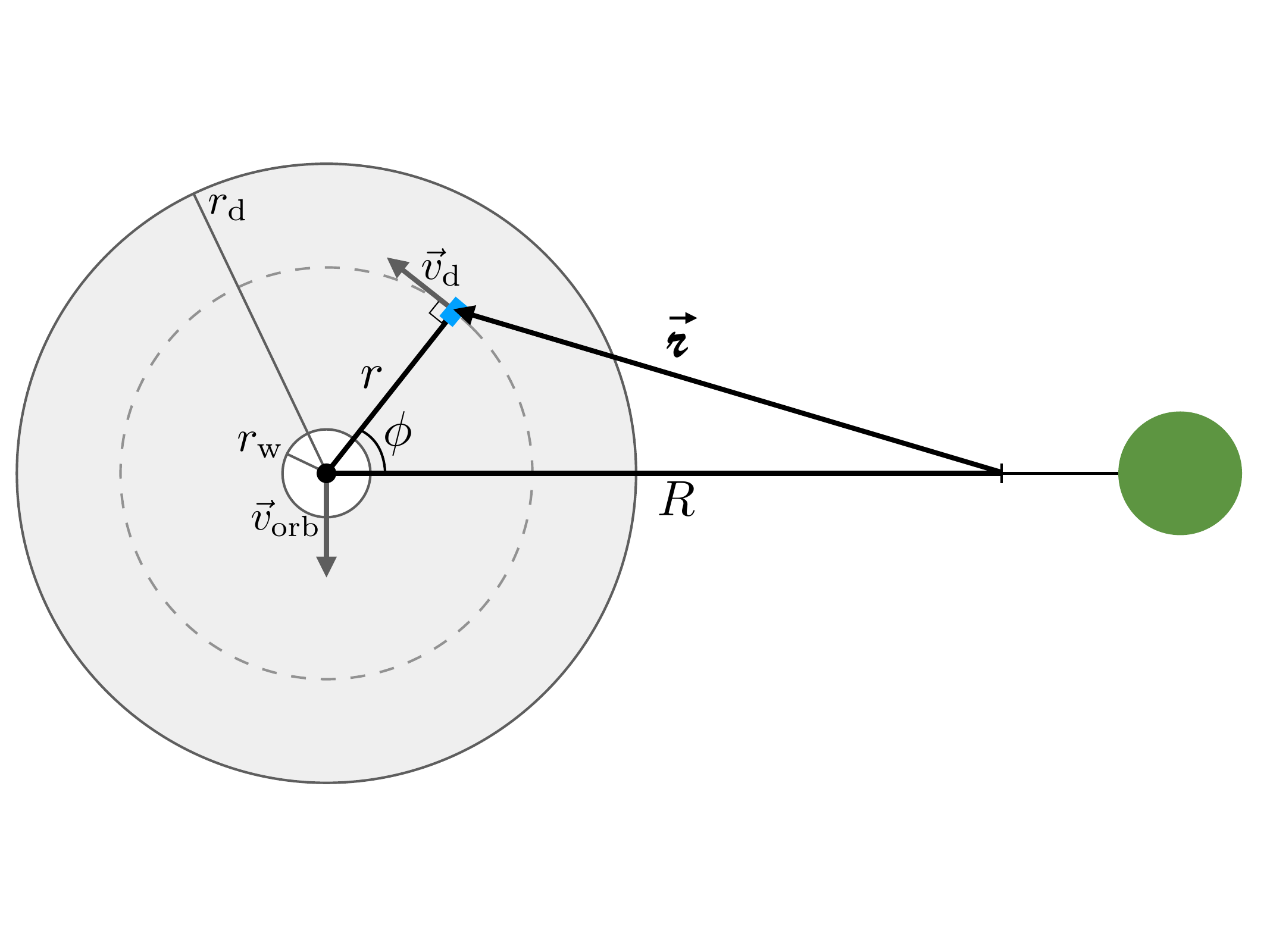}
\caption{The geometry of the accretion disk around the compact object in a binary. The gray region represents the surface of the disk wind, which extends from $r_{\mathrm{w}}$ to $r_{\mathrm{disk}}$. The dashed circle within this region represents a ring of radius $r$ within the disk.  The distance from the center of mass of the system to the accretor is $R$. The vector from center of a mass of the system to a mass element at radius $r$ and angle $\phi$ is $\scriptr{}$ is $\vec{\scriptr{}}$. 
The velocity of the mass element is the sum of the orbital velocity of the accretor $\vec{v}_{\mathrm{orb}}$ and the Keplarian velocity about the accretor at a distance $r$ within the accretion disk, $\vec{v}_{\mathrm{d}}$.
}
\label{fig:geometry}
\end{figure}

In Fig.~\ref{fig:geometry} we show the disk geometry within a binary system.
An example mass element within the disk is shown in blue. 
Each mass element in the disk wind carries away the specific angular momentum $h = \vec{\scriptr{}} \times \vec{v} $, where $\vec{\scriptr{}}$ is the vector from the center of mass of the binary to the mass element.
The velocity of the mass element about the center of mass, $\vec{v}$, is the sum of the orbital velocity of $M_2$, $\vec{v}_{\mathrm{orb}}$, and the mass element's Keplerian velocity in the disk, $\vec{v}_{\mathrm d}$, at a distance $r$ from the accretor. The specific angular momentum of each mass element is therefore

\begin{equation}
   h =  \vec{\scriptr{}} \times (\vec{v}_{\mathrm{orb}}+\vec{v}_{\mathrm{d}}).
\end{equation}   
We assume that the angular momentum of the disk and the binary are aligned.
The specific angular momentum of a mass element at radius $r$ and angle $\phi$ simplifies to, 

\begin{equation}
    h(r,\phi) = |v_{\mathrm{d}}|r + |v_{\mathrm{orb}}|R -  (|v_{\mathrm{d}}|R + |v_{\mathrm{orb}}|r)\cos\phi.
    \label{eq:specific_am_h}
\end{equation}

The total rate of angular momentum loss from all mass elements in the disk wind is,
\begin{equation}
    \dot{J}_{\mathrm{disk}} = \iint \dot{\sigma}_{\mathrm{w}}h(r,\phi) r \mathrm{d}r \mathrm{d}\phi.
\end{equation}
where the integral bounds are from 0 to $2\pi$ and from the inner radius of the disk wind, $r_{\mathrm{w}}$, to the outer radius of the disk, $r_{\mathrm{d}}$.  
Assuming $r_{\mathrm{w}}=0$ for $\beta=1$, we set our lower radial bound to 0. 
We show the rate of orbital angular momentum loss for any $\beta$ in the Appendix.
Integrating over $\phi$: 
\begin{equation}
    \dot{J}_{\mathrm{disk}} = \frac{\xi \dot{M}_{\mathrm{w}} }{r_{\mathrm{d}}^2} \int_{0}^{r_{\mathrm{d}}} \left( \frac{r}{r_{\mathrm{d}}}\right)^{\xi - 2} (v_{\mathrm d}r + v_{\mathrm {orb}}R) r \mathrm{d}r.
\end{equation}
Using $v_{\mathrm{orb}} = 2\pi R /P$, where $P$ is the orbital period, $v_{\mathrm{d}} = \sqrt{GM_2/r}$, and integrating over $r$ we find
\begin{equation}
    \dot{J}_{\mathrm{disk}} = \dot{M}_{\mathrm{w}} \left[ \frac{\xi}{\xi+{1}/{2}}  \sqrt{G M_2 r_{\mathrm{d}}} + \sqrt{ G (M_1 + M_2)\frac{R^4}{a^3} }   \right],
    \label{eq:disk_J_dot}
\end{equation}
where $a$ is the orbital separation. This is the rate of orbital angular momentum loss for a disk wind with mass loss rate $\dot{M}_{\mathrm{w}}$, mass ejection index $\xi$, and disk radius $r_{\mathrm{d}}$ for a fully non-conservative system.

We compare the orbital angular momentum loss of this disk wind to the case of isotropic re-emission, which assumes that mass ejected has the specific angular momentum of the accretor. 
Assuming a rate of mass loss of $\dot{M}_{\mathrm{w}}$,

\begin{equation}
    \dot{J}_{\mathrm{standard}} =  \dot{M}_{\mathrm{w}} \sqrt{ G (M_1 + M_2)\frac{R^4}{a^3}}.
    \label{eq:standard_J_dot}
\end{equation}
Eqn.~\eqref{eq:standard_J_dot} is the second term in the disk-wind case. Thus, the mode of mass loss via a disk wind will always remove more orbital angular momentum than the standard mode of isotropic re-emission. 

\subsection{Binary orbital evolution} \label{sec:evolve_binary}

The rate of change of a binary's separation, $\dot{a}$, is determined by the rate of orbital angular momentum loss $\dot{J}_{\mathrm{orb}}$, the donor (accretor) mass $M_1$ ($M_2$), the rate of mass loss from the donor $\dot{M}_1$, and the fraction of material removed from the system $\beta$,

\begin{equation}
    \frac{\dot{a}}{a} = -2 \left( 1 + (\beta -1) \frac{M_1}{M_2} -\frac{\beta}{2}\frac{M_1}{M}\right)\frac{\dot{M_1}}{M_1} + 2 \frac{\dot{J}_{\mathrm{orb}}}{J_{\mathrm{orb}}},
    \label{eq:a_dot}
\end{equation}
\citep[e.g.,][]{Postnov2014}. Since we are interested in the effect of varying $\dot{J}_{\mathrm{orb}}$ with different prescriptions, we do not consider any other forms of mass loss, such as stellar winds, or other forms of angular momentum loss, such as from a circumbinary disk.
We fix the rate of mass loss of the primary, $\dot{M_1}$, and vary the initial mass ratio, $q$, and mass ejection index, $\xi$. 
We compare the orbital separation when $\dot{J}_{\mathrm{orb}}$ is determined by $\dot{J}_{\mathrm{disk}}$ to when it is  determined by $\dot{J}_{\mathrm{standard}}$.
We use the fourth-order Runge-Kutta method to solve Eqn.~\eqref{eq:a_dot}.

Alternatively, we can analytically solve for the final separation $a_{\mathrm{f}}$ by replacing $\dot{J}_{\mathrm{orb}}$ with $\dot{J}_{\mathrm{standard}}$ in Eqn.~\eqref{eq:a_dot} and integrating over time. For $\beta = 1$,

\begin{equation}
    \frac{a_{\mathrm{f}}}{a_{\mathrm{i}}} = \frac{M_{\mathrm{tot, i}}}{M_{\mathrm{tot, f}}} \left(  \frac{M_{1,\mathrm{i}}}{M_{1,\mathrm{f}}}  \right)^2 \exp{ \left( -2 \frac{M_{1,\mathrm{i}}-M_{1,\mathrm{f}}}{M_2} \right)} \label{eq:a_final_standard}
\end{equation}
\citep{Postnov2014}.
Similarly, replacing $\dot{J}_{\mathrm{orb}}$ with $\dot{J}_{\mathrm{disk}}$ in Eqn.~\eqref{eq:a_dot} we find

\begin{equation}
    \begin{aligned}
    \frac{a_{\mathrm{f}}}{a_{\mathrm{i}}} = & \frac{M_{\mathrm{tot, i}}}{M_{\mathrm{tot, f}}} \left(  \frac{M_{1,\mathrm{i}}}{M_{1,\mathrm{f}}}  \right)^2 \exp{ \left( -2 \frac{M_{1,\mathrm{i}}-M_{1,\mathrm{f}}}{M_2} \right)}\\
    & \exp{\left(2\frac{\xi}{\xi+\frac{1}{2}}\int_{u_i}^{u_f} W(u)\mathrm{d}u \right)},
    \end{aligned}
    \label{eq:a_final_disk}
\end{equation}
where

\begin{equation}
    W(u) = \sqrt{\frac{1+u}{u^2}\frac{r_\mathrm{d}}{a}},
\end{equation}
 $u=1/q$ and $r_d/a$ is a function of only $q$. The only difference between this equation and Eqn.~(\ref{eq:a_final_standard}) is the supplementary exponential with $\xi$ and $W(u)$. Solutions for the integral of $W(u)$ can be found using hypergeometric functions, but they are not very enlightening. Instead, we integrate $W(u)$ numerically to cross-validate our integration procedure of the full orbital evolution equation (Eqn.~\ref{eq:a_dot}).

\section{Results} \label{sec:results}

\subsection{Dependence on $q$ and $\xi$}
To explore the differences in the two prescriptions, it is helpful to study their ratio.
Computing the ratio of $\dot{J}_{\mathrm{disk}}$ (Eqn.~\ref{eq:disk_J_dot}) to $\dot{J}_{\mathrm{standard}}$ (Eqn.~\ref{eq:standard_J_dot}) and simplifying the mass dependence in terms of the mass ratio, we find that $\dot{J}_{\mathrm{disk}}/\dot{J}_{\mathrm{standard}}$ depends only on a few variables:

\begin{equation}
    \frac{\dot{J}_{\mathrm{disk}}}{\dot{J}_{\mathrm{standard}}} = 1 + \frac{\xi}{\xi + 1/2} \sqrt{ \frac{q}{1+q} \frac{r_{\mathrm{d}}}{R} \left(\frac{a}{R}\right)^3}.
    \label{eq:j_dot_ratio}
\end{equation}
We first discuss the dependence on the term, $r_{\mathrm{d}}/R$, the ratio of the disk radius to the distance of the compact object from the CM.
As $r_{\mathrm{d}}/R$ increases, $\dot{J}_{\mathrm{disk}}$ deviates more from $\dot{J}_{\mathrm{standard}}$. 
This is expected because as $r_{\mathrm{d}}$ gets larger, mass elements that were once assumed to be near the vicinity of the accretor now extend to larger distances from the accetor and the center of mass of the binary.
This results in larger ranges for the values of $\scriptr{}$ and thus larger specific angular momentum when $\scriptr{}>R$. 
When $R$ small, the influence of a large $\scriptr{}$ is further exaggerated.
Hence, $r_{\mathrm{d}}/R$ strongly influences the importance of the accretion disk wind.

The ratio $r_{\mathrm{d}}/R$ itself simplifies to only a function of $q$, which increases with $q$.
The same is true for $(a/R)^3$ in Eqn.~\eqref{eq:j_dot_ratio}. 
As a result, we find that for equal $a$ and $\dot{M}_{\mathrm{w}}$, $\dot{J}_{\mathrm{disk}}/\dot{J}_{\mathrm{standard}}$ depends only on the mass ratio, $q$, and the mass ejection index, $\xi$.

\begin{figure}
    \centering
\includegraphics[width=0.49
\textwidth]{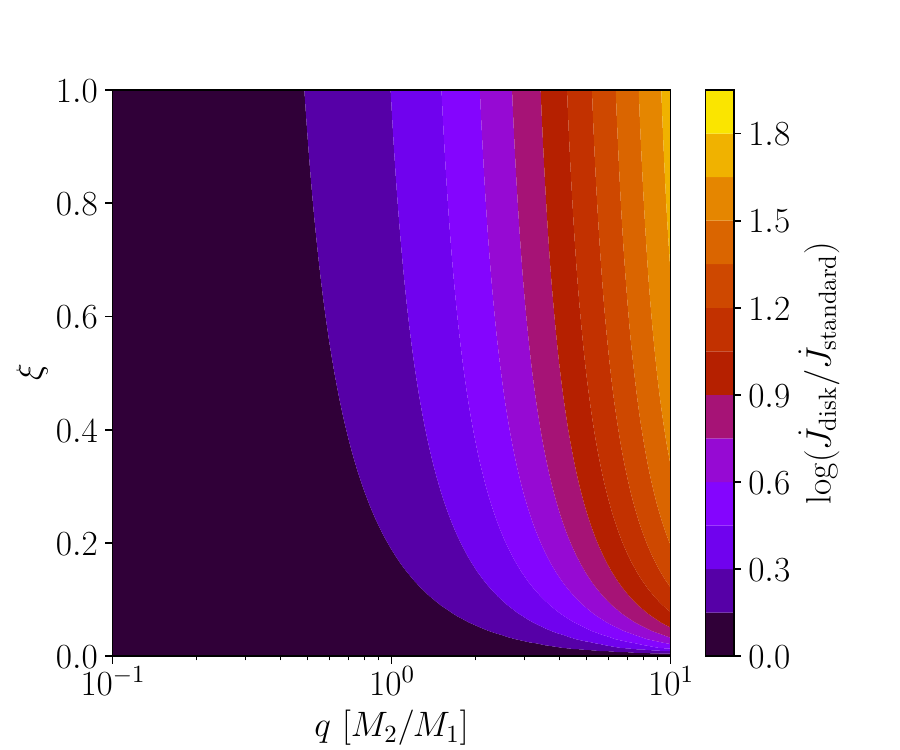}
    \caption{The ratio of the rate of angular momentum loss from a disk wind, $\dot{J}_{\mathrm{disk}}$, to angular momentum loss in the case of isotropic re-emission, $\dot{J}_{\mathrm{standard}}$ as a function of the mass ejection index, $\xi$, and mass ratio of the binary $q$. 
    This ratio is maximized for large $q$, where systems have large disk radii, and large $\xi$, where more mass loss occurs in the outer disk regions.}
    \label{fig:parameter_space}
\end{figure}

In Fig.~\ref{fig:parameter_space} we show $\log(\dot{J}_{\mathrm{disk}}/\dot{J}_{\mathrm{standard}})$ for a range of $q$ and $\xi$ values assuming a fully non-conservative model, $\beta=1$. 
The ratio $\dot{J}_{\mathrm{disk}}/\dot{J}_{\mathrm{standard}}$ increases with $q$ for all values of $\xi$ but only moderately increases with $\xi$ for a given $q$.
For equal binary separations, the disk radius increases with increasing $q$, which for the reasons listed above will increase the deviation from the standard model. 
On the other end, when $q<1$, the disk radius is small and mass loss occurs closer to the accretor, which is similar to the standard isotropic re-emission model and thus $\dot{J}_{\mathrm{disk}}/\dot{J}_{\mathrm{standard}}$ approaches 1.

The mass ejection index, $\xi$, determines how the mass loss is radially distributed on the disk surface. 
Values of $\xi$ closer to 0 describes a flat disk wind profile, while $\xi \sim 1$ describes a wind profile where more mass is ejected at the outer edges of the disk with larger $\scriptr{}$. 
As a result, $\dot{J}_{\mathrm{disk}}/\dot{J}_{\mathrm{standard}}$ increases with increasing $\xi$, but it is only apparent for large $q$ values where the disk is large. 
The highest value achieved for this range of $q$ and $\xi$ is $\simeq70$.

The values of $\dot{J}_{\mathrm{disk}}/\dot{J}_{\mathrm{standard}}$ in Fig.~\ref{fig:parameter_space} only apply at an instantaneous time and for systems with equal separations. 
We must integrate the effects of these two prescriptions through time to determine the total differences in angular momentum loss and the resulting binary separations.

\subsection{Evolution dominated by angular momentum loss or mass loss}

\begin{figure}
    \centering
    \includegraphics[width=0.49
\textwidth]{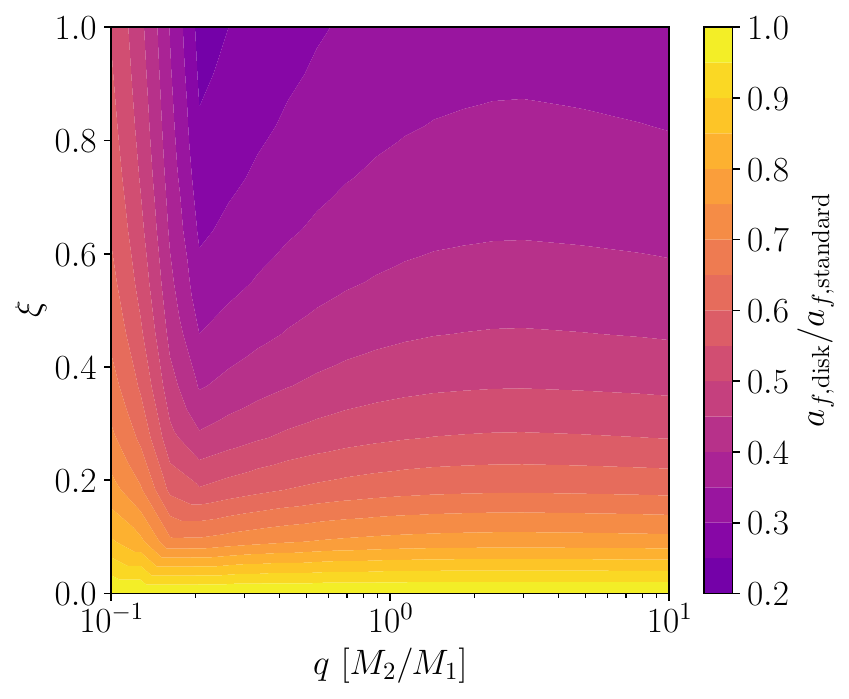}
    \caption{ Ratio of the final separation found using the disk model, $a_{f,{\mathrm{disk}}}$, to the final separation using the standard isotropic re-emission prescription, $a_{f,{\mathrm{standard}}}$, as a function of mass ejection index $\xi$ and initial mass ratio $q$.}
    \label{fig:beta1_results}
\end{figure}

In Fig.~\ref{fig:beta1_results} we show the ratio of the final separations assuming the disk-wind model, $a_{f,{\mathrm{disk}}}$, to the final separations using the standard isotropic re-emission prescription, $a_{f,{\mathrm{standard}}}$, as a function of initial mass ratio, $q$, and mass ejection index, $\xi$.
We fix $\dot{M}_1$, assume fully non-conservative MT, and integrate until the donor reaches 30\% of its initial value.
The final separations following the disk-wind model are considerably smaller, with $a_{f,{\mathrm{disk}}} \lesssim 0.5 \ a_{f,{\mathrm{standard}}}$ for most cases.
For mass ejection index $\xi \simeq 0.4$, the final separations are generally small, with $a_{f,{\mathrm{disk}}} \simeq 0.45 \ a_{f,{\mathrm{standard}}}$.
For intermediate values of $q\simeq 0.2-0.4$, the difference in final separations tends to increase.
Below these mass ratios, the separations for both $a_{f,{\mathrm{disk}}} $ and $ \ a_{f,{\mathrm{standard}}}$ reach small values, a few solar radii, and we terminate any further evolution. 

The behavior of $a_{f,{\mathrm{disk}}}/a_{f,{\mathrm{standard}}}$ with $q$ does not follow the trend for $\dot{J}_{\mathrm{disk}}/\dot{J}_{\mathrm{standard}}$ where its value is maximized at large $q$.
This can be understood by considering the geometry of the disk relative to the binary system and how the separation of the binary is affected by the physical changes to the system.

The evolution of a binary's separation, $a$, depends strongly on whether the rate of change of the separation, $\dot{a}$, is dominated by angular momentum loss or mass loss. 
In Eqn.~\eqref{eq:a_dot}, when the term associated with the donor's mass loss $\dot{M}_1$ (defined as negative) is large, the rate of change of the separation $\dot{a}$ is positive: the binary's separation increases since a loss of mass from the system decreases its binding energy, separating the two objects. 
On the other hand, when the rate of change of the separation is dominated by angular momentum loss, $\dot{a}$ is negative and the separation decreases -- removing angular momentum brings the two objects to smaller radii.
An alternate prescription to  $\dot{J}_{\mathrm{orb}}$ will have the most effect on the orbital evolution of binary systems that are dominated by angular momentum loss instead of mass loss, which, in turn, depends on the mass ratio of the binary.

For both prescriptions we consider, the mass ratio determines where (relative to the center of mass) mass loss occurs. 
With decreasing $q$, mass loss occurs farther from the center of mass of the system.
This results in larger possible values of $\scriptr{}$ with decreasing $q$, which drives a higher loss of angular momentum and thus a larger contribution to $\dot{a}$.
We therefore expect small $q$ systems to be more affected by an alternate prescription to $\dot{J}_{\mathrm{orb}}$. 
However, as discussed in the previous section, our alternate prescription  deviates less from the standard model at small $q$ values,  $\dot{J}_{\mathrm{disk}}/\dot{J}_{\mathrm{standard}}$ approaches $1$.
Integrating the differences through time for initial values of $q<0.2$ results in moderate differences. 

With increasing $q$, mass loss near the vicinity of the accretor and from a disk-wind will be closer the center of mass of the system, minimizing the angular momentum of the mass loss.
Although the rate of orbital angular momentum loss decreases, $\dot{J}_{\mathrm{disk}}/\dot{J}_{\mathrm{standard}}$ increases with $q$. 
Integrating the differences through time results in considerable differences.

At intermediate values of $q \simeq 0.2-0.4$, the ratios of the final separations reach smaller values. 
This is likely because these systems transition from low to high $q$, allowing the binary to experience the effects of high rates of angular momentum loss at low initial $q$ to the transition at high $q$ where angular momentum loss differences in the disk-wins model and standard prescription are largest.

\subsection{Varying $\beta$ and $r_{\mathrm{d}}$}
\label{sec:beta_r_disk}

In terms of $a_{f,{\mathrm{disk}}}/ a_{f,{\mathrm{standard}}}$, we do not find a significant dependence on $\beta$. At $\beta=0.5$, the difference  remains $\simeq 50\%$ for more than half of the ranges of $\xi$ and initial $q$. This is likely because although a smaller $\beta$ means less mass loss, the mass ratios throughout evolution reach higher values since now the accretor mass increases as the donor mass decreases. This results in larger disk-wind radii with large values of angular momentum loss. In this way, the dependence of $\dot{J}_{\mathrm{disk}}/\dot{J}_{\mathrm{standard}}$ on $q$ offset the decreases in mass loss. 

Varying $r_{\mathrm{d}}$ to 50\% of the Roche Lobe radius only moderately affects our results as well. This is because $\dot{J}_{\mathrm{disk}}/\dot{J}_{\mathrm{standard}}$ (Eqn.~\ref{eq:j_dot_ratio}) scales as $\sqrt{r_{\mathrm{d}}}$. As a result, the general trends are preserved and the effect of a disk wind still results in values of $a_{f,{\mathrm{disk}}} \simeq 0.5 \ a_{f,{\mathrm{standard}}}$ for $\xi \simeq 0.4$ when $r_{\mathrm{d}}$ is 50\% of the Roche Lobe radius and $\beta=1$. 
Variations to the MT efficiency and disk radius do not significantly affect the values of $a_{f,{\mathrm{disk}}}/ a_{f,{\mathrm{standard}}}$. 
 
\section{Conclusions}\label{sec:conclusions}

We have calculated the evolution 
of binary orbits using two different prescriptions for orbital angular momentum loss. 
We considered the isotropic re-emission mode where mass loss from the binary has the specific angular momentum of the accertor. 
We derived an analytic model for mass loss from the binary that is informed by observations of X-ray binary accretion disks and GRMHD simulations. 
Our analytic model accounts for angular momentum loss from a disk wind around the accretor with a mass ejection profile $ \propto r^{\xi}$, where more mass loss occurs at outer regions of the disk. 

We find that our disk-wind model results in more angular momentum loss compared to the standard case of isotropic re-emission. 
For fixed separations, the differences in angular momentum loss between the two prescriptions increase with increasing $q$.
This is because larger mass ratios have larger disks and mass loss occurs farther from the center of mass of the system, increasing the angular momentum.
For $q = 10$ and $\xi=0.4$, $\dot{J}_{\mathrm{disk}}/\dot{J}_{\mathrm{standard}}\simeq40$.
This does not translate to differences in the final orbital separations of the order of 40.
For fixed orbital separations, the effect of orbital angular momentum loss (due to ejection of mass around an accretor) on the orbital separation decreases with increasing $q$. 
Thus there is a counteracting dependence on $q$ that reduces the impact of the differences in angular momentum loss between the two prescriptions on the orbital separation. 
\emph{Nonetheless, the orbital separation of a binary can be considerably affected by mass loss via a disk wind.}

Even though we focused our discussion on magnetically driven winds, our model can also apply to radiatively driven winds if the wind density profile roughly follows a power law. Power law behavior for the density profile of the radiatively driven wind is observed in \cite{tomaru_what_2023}. We calculate a mass ejection index of $\xi\simeq0.5$ from their density profile, consistent with our parameter space exploration.

Here we maximized the effect of a disk wind on the orbital evolution so that we may assess its extreme behavior. 
For example, we fix the MT rate throughout the simulation until the donor reaches 30\% of its initial value, which is a large oversimplification.
Similarly, we do not consider the possibility of the MT becoming unstable.

An alternative model for orbital angular momentum loss will have consequences for the stability of MT.
Using \texttt{COMPAS}, \cite{Willcox2023} show how the fraction of systems that undergo stable MT strongly depends on changes to the orbital angular momentum loss alone.
Their results provide clear motivation for determining a physically-motivated model for the angular momentum loss during stable MT. 

The presence of a disk wind is supported by observations and our calculations show that it can have an important effect on the final orbital separation of a binary system.
To fully determine the impact on the stability of MT and the outcome of binary evolution, a large parameter study across a range of donor masses, mass ratios, and initial orbital periods is needed.
Determining the duration of these MT phases is crucial as that would determine the extent to which the angular momentum loss from the disk wind affects the binary orbit. 
A careful consideration of realistic MT phases, and appropriate duration of MT phases, can be carried out with detailed stellar and binary evolutionary codes such as \texttt{MESA} \citep{Paxton2011, Paxton2013,Paxton2015,Paxton2019}.
We leave this work for a future study.

\begin{acknowledgements}

The authors thank Sasha Tchekhovskoy, Enrico Ramirez-Ruiz, and Beverley Lowell for helpful discussions; Christopher Berry and Zoheyr Doctor for comments on the manuscript.
M.G.-G. is supported by Northwestern University.
JJ acknowledges support by the NSF AST-2009884 and NASA 80NSSC21K1746 grants.
V.K. is supported by the Gordon and Betty Moore Foundation through grant GBMF8477 and by Northwestern University.

\end{acknowledgements}

\software{\citep{townsend2019}
\texttt{Matplotlib} \citep{Hunter2007}; 
\texttt{NumPy} \citep{vanderwalt2011};
\texttt{Pandas} \citep{mckinney-proc-scipy-2010}.
}

\bibliography{ms}
\bibliographystyle{aasjournal}

\appendix

Here we provide more details for the calculation of the specific angular momentum. Then we derive $\dot{J}_{\mathrm{disk}}$ for a partially conservative MT system, $\beta<1$.
An illustration of the geometry and variables are provided in Fig.~\ref{fig:geometry}

As described in Section~\ref{sec:angular_momentum_loss}, the specific angular momentum of a mass particle within the disk is the sum of velocities and its distance from the center of mass.

\begin{equation}
   h = (\vec{v}_{\mathrm{orb}}+\vec{v}_{\mathrm{d}}) \times \vec{\scriptr{}}.
   \label{eq:specific_am_h_appendix}
\end{equation}   
The vector to the center of mass, $\vec{\scriptr{}}$, is given by, 

\begin{equation}
   \vec{\scriptr{}} = (R - r\cos{\phi}) \hat{x} + (r \sin{\phi})\hat{y},
\end{equation}   
and the velocity vectors are given by, 

\begin{equation}
    \vec{v}_{\mathrm{orb}} = -|v_{\mathrm{orb}}| \hat{y}
\end{equation}
and

\begin{equation}
    \vec{v}_{\mathrm{d}} = (|v_{\mathrm{d}}| \sin{\phi}) \hat{x} + (|v_{\mathrm{d}}| \cos{\phi}) \hat{y}.
\end{equation}
The cross product simplifies to,

\begin{equation}
    h(r,\phi) = |v_{\mathrm{d}}|r + |v_{\mathrm{orb}}|R -  (|v_{\mathrm{d}}|R + |v_{\mathrm{orb}}|r)\cos\phi.
    \label{eq:specific_am_h_final_appendix}
\end{equation}
Given Eqn.~\ref{eq:specific_am_h_final_appendix} and $\dot{\sigma}_{\mathrm{w}}$, we integrate over the surface of the disk to calculate the total angular momentum loss, 

\begin{equation}
    \dot{J}_{\mathrm{disk}} = \iint \dot{\sigma}_{\mathrm{w}}h(r,\phi) r dr d\phi.
\end{equation}

In this case of a partially conservative MT system, we assume the rate of wind mass loss per unit area, $\dot{\sigma}_{\mathrm{w}}$, follows the same power law as in the fully non-conservative case. 
The accretion rate onto the compact object, and total mass loss from the disk is therefore controlled by adjusting the inner disk radius to $r_{\mathrm{w}}$. A description of $r_{\mathrm{w}}$ was provided in Section~\ref{sec:mass_loss}. The presence of $r_{\rm w}$ changes the normalization of the wind mass loss per unit area to be
\begin{equation}
    \dot{\sigma}_{\rm w} =  \frac{\xi \dot{M}_w}{2\pi r_d^2}\frac{1}{1 - \Tfrac{r_w}{r_d}^{\xi}} \left(\frac{r}{r_{\rm d}}\right)^{\xi-2}.
    \label{eq:sigma_dot_wind_r}
\end{equation}

Using $v_{\mathrm{d}} = \sqrt{GM_{2}/r }$ and integrating over $d\phi$,

\begin{equation}
    \dot{J}_{\mathrm{disk}} =  \frac{\xi\dot{M}_{\mathrm{w}}}{ r_{\mathrm{d}}^2- r_{\mathrm{d}}^2\Tfrac{r_w}{r_d}^{\xi}} \int_{r_{\mathrm{w}}}^{r_{\mathrm{d}}} \left( \frac{r}{r_{\mathrm{d}}}\right)^{\xi - 2} \left(r\sqrt{G M_2 /r} + v_{\mathrm {orb}}R\right) r dr,
\end{equation}
Rearranging $r_{\mathrm{d}}$ to facilitate an integral over $dr/r_{\mathrm{d}}$,
\begin{equation}
    \dot{J}_{\mathrm{disk}} = \frac{\xi\dot{M}_{\mathrm{w}} }{1 - \Tfrac{r_w}{r_d}^{\xi}} \int_{r_{\mathrm{w}}}^{r_{\mathrm{d}}} \left( \frac{r}{r_{\mathrm{d}}}\right)^{\xi - 1} \left(\sqrt{G M_2 r_{\mathrm{d}}} \left(\frac{r}{r_{\mathrm{d}}}\right)^{1/2} + v_{\mathrm {orb}}R\right) \frac{dr}{r_{\mathrm{d}}},
\end{equation}
This is now a simple integral over $dr/r_{\mathrm{d}}$ which simplifies to
\begin{equation}
  \dot{J}_{\mathrm{disk}} = 
 \dot{M}_{\mathrm{w}} \left[v_{\rm orb}R+\frac{\xi}{\xi+\frac{1}{2}}\sqrt{GM_{2}r_d}\,\,\frac{1 - \Tfrac{r_w}{r_d}^{\xi+\frac{1}{2}}}{1 - \Tfrac{r_w}{r_d}^{\xi}}\right],
\end{equation}
 where $r_{\mathrm{w}} = r_{\mathrm{d}}(1-\beta)^{1/\xi}$.
 This reduces to Eqn.~\ref{eq:disk_J_dot} when $\beta=1$.

\end{document}